\documentclass{acm_proc_article-sp}
\usepackage{graphicx}
\usepackage[flushleft]{threeparttable}
\usepackage{booktabs} 
\usepackage{amsmath,varwidth,array,ragged2e}

\begin{document}

\conferenceinfo{}{Bloomberg Data for Good Exchange 2016, NY, USA}

\title{Measuring the Impact of Urban Street Trees on Air Quality and Respiratory Illness}
\subtitle{A Data-Driven Approach to Environmental Justice}

\numberofauthors{3}
\author{
\alignauthor
Yuan Lai\\
       \affaddr{Dept. of Civil and Urban Engineering \& Center for Urban Science and Progress, New York University}\\
       \affaddr{New York, NY}\\
       \email{yuan.lai@nyu.edu}
\alignauthor
Constantine E. Kontokosta, PhD\\
       \affaddr{Dept. of Civil and Urban Engineering \& Center for Urban Science and Progress, New York University}\\
       \affaddr{New York, NY}\\
       \email{ckontokosta@nyu.edu}
}

\maketitle
\begin{abstract}
New streams of data enable us to associate physical objects with rich multi-dimensional data on the urban environment. This study presents how open data integration can contribute to deeper insights into urban ecology. We analyze street trees in New York City (NYC) with cross-domain data integration methods by combining crowd-sourced tree census data - which includes geolocation, species, size, and condition of each street tree - with pollen activity and allergen severity, neighborhood demographics, and spatial-temporal data on tree condition from NYC 311 complaints. We further integrate historical data on neighborhood asthma hospitalization rates by Zip Code and in-situ air quality monitoring data (PM 2.5) to investigate how street trees impact local air quality and the prevalence of respiratory illnesses. The results indicate although the number of trees contributes to better air quality, species with severe allergens may increase local asthma hospitalization rates in vulnerable populations. 
\end{abstract}





\section{Introduction}
Environmental justice is the fair treatment and meaningful participation of all people, regardless of their ethnic or socio-economic background, in the formulation of public policy and environmental regulations \cite{epa}. As increasing population density, high living costs, and climate change have created significant urban challenges, underprivileged communities often face more severe quality--of-life conditions with fewer resources and limited access to public services. In most global cities, including New York City (NYC), there are increasing concerns about air quality and respiratory illness among low-income neighborhoods, immigrants, and ethnic-minority groups due to ambient air pollution, housing conditions, or a lack of awareness of, or access to, basic health-care \cite{Columbia}. The multidimensional environmental, demographic, social, and operational factors involved make environmental justice a complex issue requiring collaborative efforts from multiple stakeholders.

A large number of urban infrastructure objects have been digitalized for agency operations and civic engagement at the community, district, and city level. Publicly-available urban data platforms are an important part of citizen science, information democratization, and analytics-supported operational decisions. However, it is challenging to translate individual datasets into meaningful information, due to the absence of context, real-time situational factors, or social sentiment. Administrative data are typically generated in siloes within a respective agency, and structured for specific domains (transportation, environment, land use, etc.) and uses without considering future data integration opportunities. The absence of readily-scalable approaches for integrating and localizing urban open data further exacerbates the digital divide that is particularly pronounced in low-income communities. Thus, methods to transform urban data into local insights for community decision-making is an urgent issue in the growing field of civic analytics \cite{kontokosta2016quantified}.

Urban infrastructure and public facilities, such as street trees, light poles, parking meters, or bicycle racks, are physical objects located at a fixed location. Digitally, such objects often represent as points with a unique ID, status, and geo-location, collected by city agencies or volunteers. Street trees are critical ecological infrastructure for cities, given their role in mitigating climate change, positive impacts on promoting active living, and aesthetic contributions to property values. In NYC, there are 652,169 documented street trees contributing an estimated total annual benefit of \$122 million \cite{treefacts}. Growing concerns around urban climate and quality-of-life necessitate multi-disciplinary research on how we plant and manage street trees as part of urban green infrastructure \cite{locke2010prioritizing}.

It is still largely unknown exactly how urban forestry impact local air quality and public health based on tree canopy density and species. Although street trees have been proven to contribute to a lower prevalence of early childhood asthma, certain species are considered as potential source of allergen to exacerbate atopic asthma \cite{lovasi2008children}. A previous study on multiple cities in Canada finds tree pollen as a significant cause of asthma and allergic sensitization for clinical visits \cite{dales2008tree}. Another study in NYC demonstrates a significant correlation between tree pollen peak season and allergy medication sales by borough \cite{sheffield2011association}. Spatial patterns of street trees further reveal issues in environmental justice and health disparities by neighborhood in NYC. A previous study on asthma hazards in Greenpoint/Williamsburg, a neighborhood in Brooklyn, NY, shows that low-income communities and minority groups face higher exposure to air pollution resulting from lower tree canopy coverage \cite{corburn2002combining}. These studies reinforce the importance of studying street trees in urban environments, but the findings are constrained by data and methodological limitations, as most rely on few monitoring sites or district-level statistics.  

In this study, we present a data-driven approach to measure the localized environmental health impact of street trees in NYC. We begin by integrating multiple datasets and conduct a spatial join to the point locations of over 600,000 street trees. We then analyze the correlation between the number, size, and species of tree with air quality (particulate matter) data from the NYC Department of Environmental Protection and asthma hospitalization rates at the neighborhood level from the NYS Department of Health. We conclude with a discussion of our findings and its applications, limitations, and future work.

\section{Methodology}
The increasing availability of municipal open data creates a rich resource for data-supported urban operations, but require proper data cleaning and computing techniques for actionable results. We contextualize street trees through a cross-domain data integration approach. Each tree has time-invariant features from the tree census, including its location and species. By combining domain knowledge in plant taxonomy, we can translate tree count data with more meaningful information, such as pollen offenders by species, blooming period by seasons, and toxic species. 

Beyond fixed attributes, we further infer trees' condition based on surrounding events. We collect extensive data on street trees, local 311 complaints, population, land use, air quality, and respiratory illness (Table~\ref{table:data}). We associate each tree with local 311 complaints to understand public observation and reporting of tree health. We develop a scalable model providing insights at both the community level and based on a specific location. Finally, we integrate air quality and asthma hospitalization rate data to measure the environmental and public health impacts of urban street trees at the neighborhood level.

\begin{table*}
\leavevmode \caption{Data Collection}
\label{table:data}
\footnotesize
\centering
\renewcommand{\arraystretch}{1}
\begin{tabular}{l l l l l }
\hline
\ Data & Source & Period & Spatial Unit\\
\ Tree Census & Dept. of Parks and Rec. & 2015 & Geo-point\\
\ Population Census & U.S. Census & 2010 & NTA*\\
\ Complaints on Trees & NYC 311 & 2010-present & Geo-point\\
\ Asthma Hospitalization & N.Y. Dept.of Health & 2012-2014 & Zip Code\\
\ Community Air Survey & NYC Dept.of Health & 2015 & UHF*\\
\ Air Survey Monitors & NYC Dept.of Health & 2008-2013 & Geo-point\\
\ Land Use (PLUTO*) & NYC Dept.of City Planning & 2009-2016 & Tax Lot\\
\hline
\end{tabular}\\
*NTA: Neighborhood Tabulation Area;\\
*UHF: United Hospital Funds Neighborhoods;\\
*PLUTO: Primary Land Use Tax Lot Output.\\
\end{table*}

\subsection{Data Collection}
In 2015, the Department of Parks and Recreation initiated the third street tree census, which became the largest participatory urban forestry project in U.S. history. From 2015 to 2016, the project involved more than 2,240 volunteers to map 666,134 street trees citywide \cite{nyctrees}. The final data were published on the NYC open data platform for research and public interest. The NYC tree census dataset (cleaned \textit{n}=652,169) provide each street tree's location (latitude and longitude), species, diameter at breast height (dbh), surrounding sidewalk condition (during survey), Neighborhood Tabulation Area (NTA), and Zip Code. 

In addition to tree census data, we extract data from NYC 311, a non-emergency municipal service request system, which receives more than 60,000 complaints annually (2010 to present) on dead or damaged trees, overgrown branches, and requests for new trees from local citizens. Each complaint is reported as a geo-located incident with time-stamp, providing a unique source of near-real-time information on citizens' interaction with trees. To capture local reporting about street trees, we query from NYC 311 data by complaint category from 2010 to present (\textit{n}=463,376), including local complaints on dead or damaged trees, and service requests for new trees.  

We collect asthma hospital discharges at the Zip Code level (\textit{n}=168) and NYC Community Air Survey data by United Hospital Funds (UHF) districts (\textit{n}=43) to investigate how street trees relate to local air quality and respiratory illness rates. In order to gain insights at high-spatial resolution, we also utilize air quality monitoring data from 2008-2013 reporting on local PM 2.5 concentrations with specific sensor geolocation (\textit{n}=162) and observation time (four seasons per year). We use population data from the U.S. Census reported for each NTA (\textit{n}=195), including total population and population by age groups. Since land use and building density may have an impact on air quality, we integrate tax lot data from the Department of City Planning's Primary Land Use Tax Lot Output (PLUTO) database to quantify building density, land use, and building space usage surrounding each air quality sensor location.

\subsection{Contextualizing with Domain Knowledge}
Domain knowledge is key enabler to contextualize urban data for meaningful insights. We associate pollen activity and severity based on tree species to enrich the informational value of tree census data. We focus on pollen due to increasing concerns about seasonal allergies and asthma caused by pollen allergen. We construct a dataset on tree pollen attributes by species through a literature review and on-line research (Table~\ref{table:pollen}). We create an index score to measure pollen impact by combining a ratio of vulnerable population (age <14 or > 60) and severe allergen ratio as (Eq.~\ref{eq:score}):

\begin{equation}\label{eq:score}
\begin{split}
Pollen\,Impact_{i} = (\frac{\sum Trees\,with\,Severe\,Allergen}{\sum Trees})_{i} \\ 
\times (\frac{\sum Vulnerable Population}{\sum Population})_{i}
\end{split} \end{equation}


\begin{table*}[ht]
\leavevmode \caption{Tree taxonomy as object attributes. }
\label{table:pollen}
\footnotesize
\setlength{\tabcolsep}{1.5pt}
\centering
\renewcommand{\arraystretch}{1}
\begin{tabular}{l l l l}
\hline
\ Tree Species & Allergic Pollen & Allergen Severity& Active Season\\
\ String & Binary (1/0) & Categorical (high/moderate/low) & Categorical\\
\hline
\end{tabular} \end{table*}

By combining tree census data with domain knowledge in plant taxonomy, the resultant dataset can serve local residents through a mapping dashboard and location-based mobile applications, so the general public can be informed about the prevalence of street trees with active allergens during each season.

\subsection{Integrating Real-time Situational Information}
To further enrich our analysis, we integrate real-time situational information from NYC 311 complaints on trees. Each year there are about 60,000 complaints related to trees, providing dynamic information on how the local population observes and engages with urban ecology during different seasons or extreme weather events. We first query a subset of NYC 311 complaint data (2010-present) by complaint category related to trees. Each complaint was documented as a geo-point with caller's location (latitude, longitude), a time-stamp for reporting time, complaint type, and zip code. We then spatially join each complaint with its neighborhood boundary defined by NTA, and associate neighborhood demographic attributes to each complaint. 

Although most complaints report on specific trees (damaged, dead, or overgrown), they capture the geo-location of the caller instead of a specific tree. To associate such information back to trees, we create a spatial query algorithm to extract surrounding complaints for each tree. In the Python language environment, we generate a 100-meter radius buffer for each tree's geo-location, and extract 311 complaints on trees that occurred within the buffer from 2010 to present. In this way, we estimate each tree's condition through inferring local complaints by spatial proximity. 

We use asthma hospital discharge data by Zip Code to investigate potential correlation between street trees and asthma rate. We use air quality survey data by UHF district to measure how tree density and species may have impact on local air quality (based on PM 2.5 levels). For in-situ air monitoring data, we use a similar spatial query approach to extract the total number of trees, total number of species, and tree counts by species within 100-meter buffer for each air quality sensing location. Finally, we use regression models to investigate street trees' impact on local air quality and asthma hospitalization rates.

\section{Findings}
\begin{figure}[!ht]
\centering
\includegraphics[width=.45\textwidth]{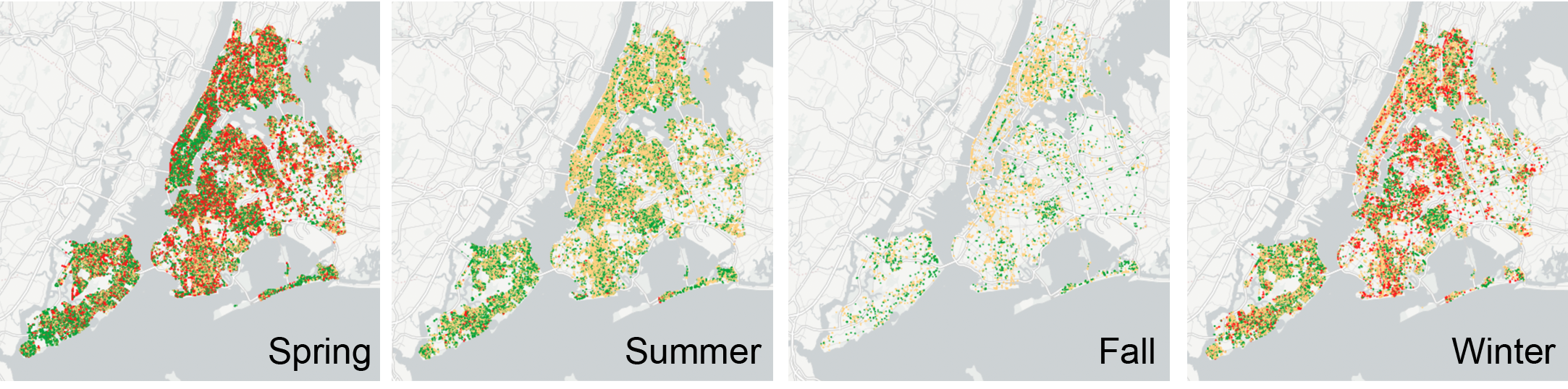}
\caption{Street tree pollen activity mapped by season.}
\label{fig:pollen}
\end{figure}

\begin{figure}[!ht]
\centering
\includegraphics[width=.45\textwidth]{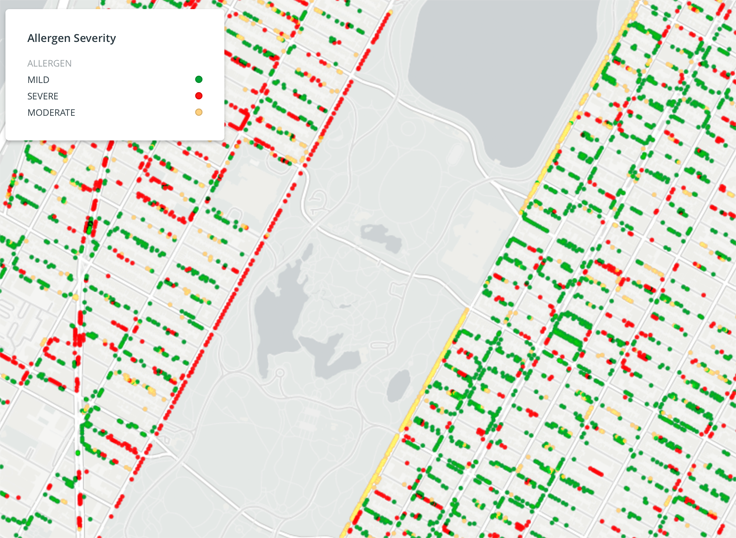}
\caption{Pollen allergen activity by tree species during spring (Central Park).}
\label{fig:central_park}
\end{figure}

\begin{figure*}[!ht]
\centering
\includegraphics[width=.8\textwidth]{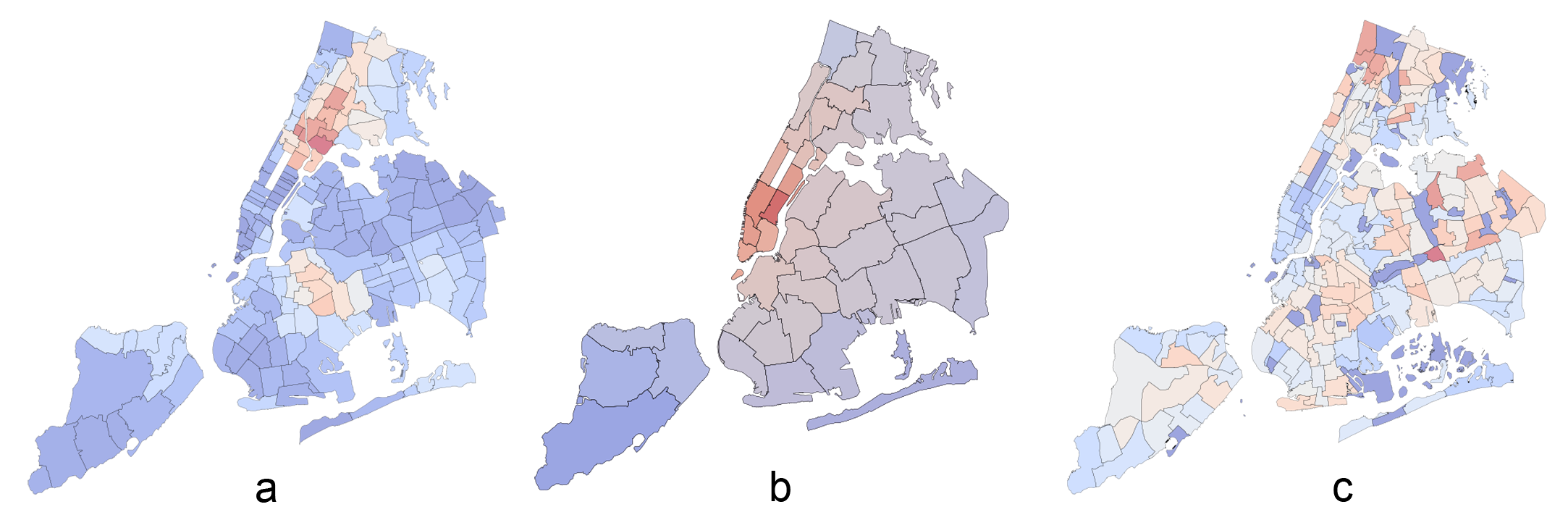}
\caption{ a. Asthma ED visit rate (per 10,000 population) by hospital discharges at Zip Code level; b. Community Air Survey on air quality (PM 2.5) by United Hospital Foundation (UHF) district in 2015; c. Street tree pollen impact by allergen severity and vulnerable population per Neighborhood Tabulation Area.}
\label{fig:pollen_pop2}
\end{figure*}

\begin{figure}[!ht]
\centering
\includegraphics[width=.4\textwidth]{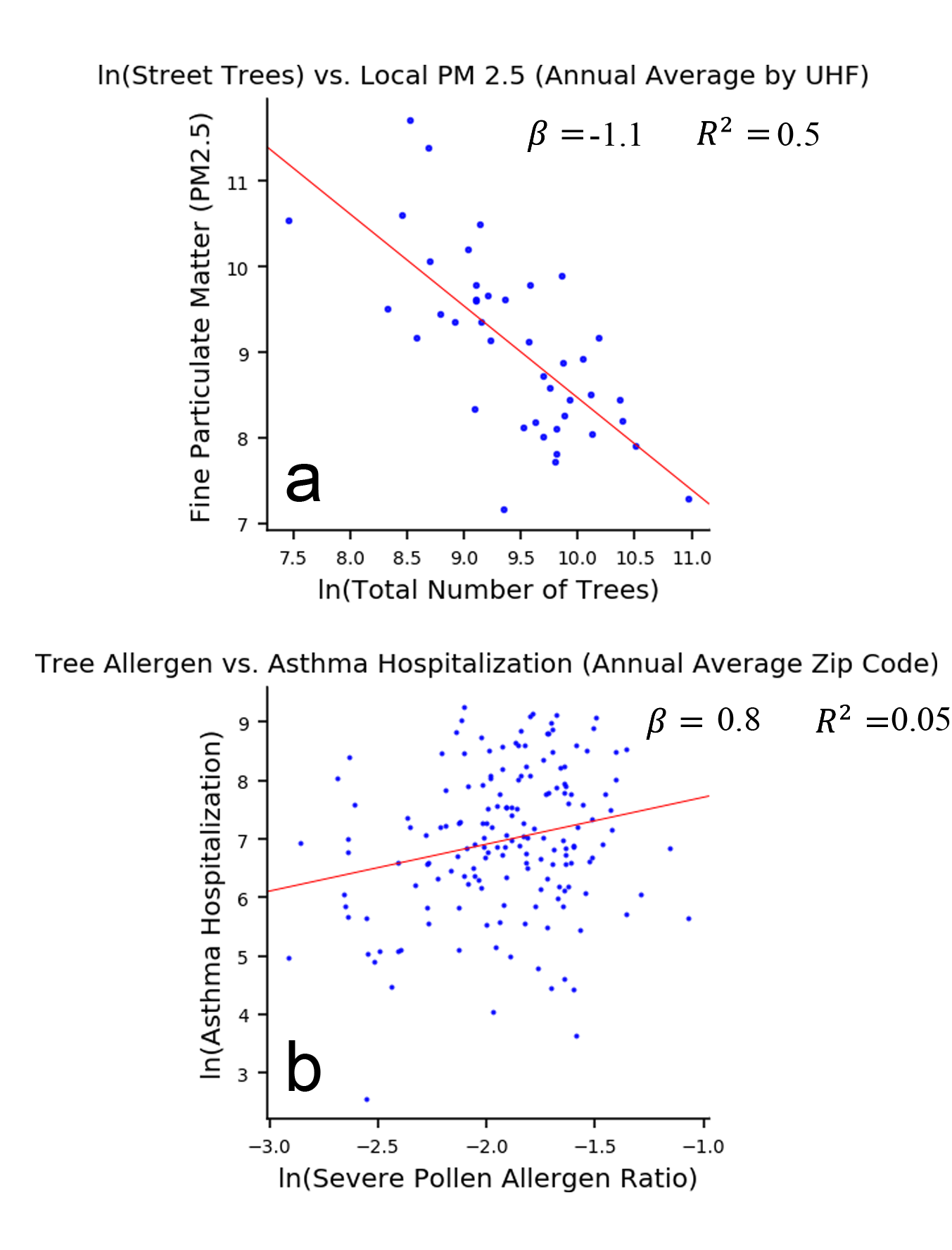}
\caption{Regression modeling results on street tree, air quality, and asthma rate.}
\label{fig:regression}
\end{figure}

A comparison of spatial patterns of asthma ED visits by discharge Zip Code, local air quality survey data of PM 2.5 levels, and pollen impact reveals the complex relationship between ambient air quality, respiratory illness, local population, and neighborhood environment (Figure~\ref{fig:pollen} \& ~\ref{fig:central_park}). The spatial disparity between asthma rate (Figure~\ref{fig:pollen_pop2}a) and PM 2.5 levels (Figure~\ref{fig:pollen_pop2}b) indicates potential confounding factors in the prevalence of respiratory illness besides air quality, such as local trees and population, as one would expect (Figure~\ref{fig:pollen_pop2}c). We run OLS linear regression models to investigate how local street trees may impact air quality and asthma hospitalizations. The results indicate an overall benefit of street trees on local air quality (Figure~\ref{fig:regression}a).

\begin{table}[ht]
\leavevmode \caption{Multivariate OLS regression results on tree species and asthma hospitalization by Zip Code Area.}
\label{table:regression}
\centering
\footnotesize
\setlength{\tabcolsep}{5pt}
\renewcommand{\arraystretch}{1.2}
\begin{threeparttable}
\begin{tabular}{l l l l l}
\hline
\multicolumn{4}{l}{Dependent Variable = $ln$(Asthma ED Visit)}\\
\hline
\ Model Variable & &Coeff. &(Std. Err.)& \\
\ $ln$(Total Trees Count)& & -0.38**  & 0.18& \\
\ $ln$(American Linden)& & -0.46***  & 0.13 &\\
\ $ln$(Callery Pear)& & -0.38***  & 0.08 &\\
\ $ln$(American Elm)& & 0.19**  & 0.08 &\\
\ $ln$(Japanese Zelkova)& & 0.31***  & 0.09 &\\
\ $ln$(Little Leaf Linden)& & 0.33***  & 0.09 &\\
\ $ln$(Honey Locust)& & 0.84***  & 0.14 &\\
\hline
\ Sample size ($N$)&  & 174 &&\\
\ Adjusted $R^2$ &  & .564  &&\\
\ $F$-test & & 25.85***  &&\\
\hline
\end{tabular} 
\begin{tablenotes} \item[ ] NOTE: Coeff.= coefficient and Std. Err.= standard error. *** = significant at 99\% ($p\leq 0.01$); **=significant at 95\% ($p\leq 0.05$); *=significant at 90\% ($p\leq 0.10$). Model variables are defined in Table 3. \end{tablenotes} \end{threeparttable}
\end{table}


Although the initial model indicates there is no significant correlation ($r^2=0.05$) between tree allergen ratio and asthma hospitalization rate (Figure~\ref{fig:regression}b), we further aggregate street trees by species at the Zip Code level to explore how specific species may impact localized asthma rates. We select each species with a median total number larger than 20 trees per zip code, and run a multivariate linear regression model to measure their impact on local asthma rates. The results reveal a profound effect by tree species. Our findings reveal the overall density of street trees contributes to a lower number of asthma ER visits, but the impact varies by tree species (Table~\ref{table:regression}). Certain species, such as Honey Locust and Little Leaf Linden, have significant positive correlations with local asthma hospitalization rates. We also find similar results when aggregating data by tree genera. We run a OLS linear panel regression model on in-situ air quality sensing data (162 locations, 4 seasons in 5 years) with surrounding trees and building density, while holding seasonality as a fixed effect. The result has a relatively low R-squared value ($r^2$=0.22), indicating unobserved or confounding factors that further explain the relationship.

\begin{figure}[!ht]
\centering
\includegraphics[width=.45\textwidth]{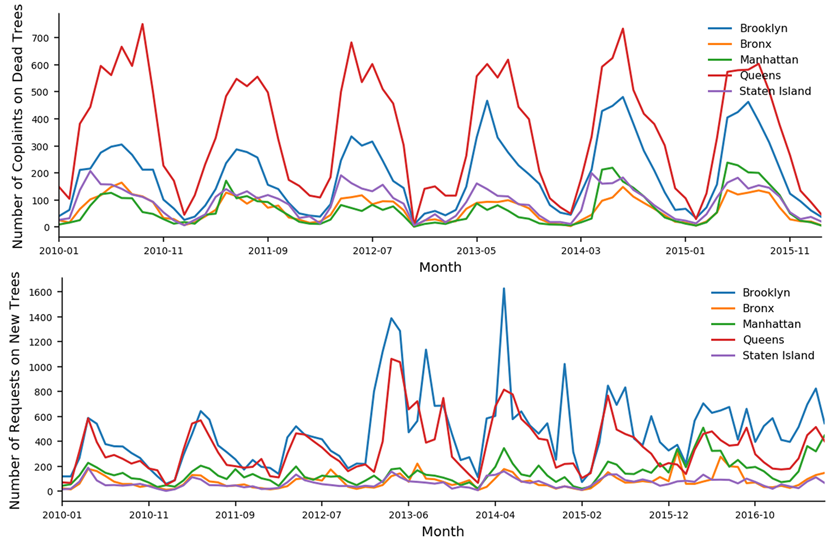}
\caption{NYC 311 Complaints on dead trees and requests for new trees by borough.}
\label{fig:complaint}
\end{figure}

Integrating tree data with local complaints reflects how residents interact with street trees. A comparison by complaint category and borough reveals different public awareness and community engagement across the City. For instance, although residents from Queens made more complaints about dead trees than those in Brooklyn, they also have fewer requests for new trees (Figure~\ref{fig:complaint}). This discrepancy may indicate a lack of knowledge about possible city services (e.g. the ability to request a new tree) within certain neighborhoods. Local complaint data indicate a regular seasonal pattern of local engagement with trees, possibly due to tree growth, weather events, and outdoor activity intensity. Using our spatial query algorithm, we associate each tree with its surrounding complaints, to further investigate spatial, temporal, and typological (tree species) patterns for predictive modeling.

\section{Discussion}
Neighborhood health disparities and environmental justice are complex issues involving environmental factors, demographics, housing, transportation, and public services \cite{locke2010prioritizing}. This study contributes to a comprehensive data integration approach for analyzing local environmental health conditions. By contextualizing tree census data with domain knowledge and local situational information, we build a robust model for evaluating the potential public health benefits of urban street trees. For instance, the New York City Community Health Profiles, a comprehensive neighborhood health report by Department of Health and Mental Hygiene in 2015 indicates Bronx Community District 1 (Mott Haven and Melrose) has the highest child asthma hospitalization rate \cite{health_profile}. Our approach can further infer local conditions on tree species and surrounding complaints, providing additional context for health data. Of course, urban forestry is just one of many factors that influences local air quality and environmental health. Again, taking Bronx Community District 1 as an example, besides its poor air quality (PM 2.5 levels of 10.0 micro-grams per cubic meter, compared with 9.1 in the Bronx and 8.6 citywide), the neighborhood also has one of the highest rates of housing maintenance defects (79\%, and the 2nd worst conditions citywide), which is another cause of respiratory illness \cite{health_profile}. Thus, further investigations in environmental health requires more extensive data integration including land use, transportation, energy usage, and housing conditions \cite{jain2014big}.


Our analysis is limited by the absence of a comprehensive plant taxonomy database on pollen activity and allergen severity. Also, since the NYC tree census data only counts street trees, there are a large number of trees in parks and open spaces not captured in the public data \cite{zandbergen2009methodological}. Thus, one further expansion of this work is to integrate trees in parks and open spaces in the City for a more complete evaluation of the impact of urban trees. 

\section{Conclusion}
In this exploratory research, we provide a data-driven approach to measure the impact of urban street trees on air quality and respiratory illness. We illustrate how cross-domain data integration can address complex environmental justice issues by quantifying local environmental, demographic, and socio-economic characteristics. Results indicate that although street trees contribute to better air quality, certain species may be a local source of allergens that can trigger or exacerbate underlying asthma conditions. Spatial disparities between air quality, asthma rates, and tree pollen impact indicate unobserved factors in neighborhood environmental health. Despite the limitations, this study provides a model for creating more meaningful insights from urban data relating to ecology and public health. Localized urban data provide community-based knowledge for residents and encourage public engagement in participatory urban sensing or citizen science projects that can raise awareness of public health, environmental justice, and access to municipal services \cite{kontokosta2016quantified,kontokosta2016}. This process requires the collective efforts of city agencies, domain experts, data scientists, and local communities. 

\nocite{*}
\bibliographystyle{abbrv}
\bibliography{main}
\end{document}